\documentclass{article}

\usepackage{arxiv}

\usepackage[utf8]{inputenc} % allow utf-8 input
\usepackage[T1]{fontenc}    % use 8-bit T1 fonts
\usepackage{hyperref}       % hyperlinks
\usepackage{url}            % simple URL typesetting
\usepackage{booktabs}       % professional-quality tables
\usepackage{amsfonts}       % blackboard math symbols
\usepackage{nicefrac}       % compact symbols for 1/2, etc.
\usepackage{microtype}      % microtypography
\usepackage{lipsum}		% Can be removed after putting your text content
\usepackage{graphicx}
\usepackage{natbib}
\usepackage{doi}

\title{ Integrated dataset for air travel and reported Zika virus cases in Colombia (Data and Resources Paper)}

\author{ 
	Aiman ~Soliman, Priyam ~Mazumdar, Aaron ~Hoyle-Katz  \\
	National Center for Supercomputing Applications\\
        University of Illinois Urbana - Champaign\\
	Urbana, IL 61801 \\
	\texttt{\{asoliman, priyamm2, aaronh2\}@illinois.edu} \\
        \AND
 	Brian ~Allan \\
	School of Integrative Biology\\
        University of Illinois Urbana - Champaign\\
	Urbana, IL 61801 \\
	\texttt{ballan@illinois.edu} \\
        \AND
 	Allison	~Gardner \\
	School of Biology and Ecology\\
        University of Maine\\
	Orono, ME 04469 \\
	\texttt{allison.gardner@maine.edu} \\
}
\renewcommand{\shorttitle}{Integrated Dataset for Air Travel and Zika Virus Cases}

\hypersetup{
pdftitle={A template for the arxiv style},
pdfsubject={q-bio.NC, q-bio.QM},
pdfauthor={Aiman ~Soliman, Priyam ~Mazumdar, Aaron ~Hoyle-Katz, Brian ~Allan, Allison ~Gardner},
pdfkeywords={Human Mobility, Open Access Data, Network Analysis, Zika Virus, Air Travel, Colombia, Epidemiology}}

\begin{document}
\maketitle

\begin{abstract}
This open-access dataset provides consistent records of air travel volumes between 205 airport catchments in Colombia and the associated number of reported human cases of Zika virus within these catchments during the arbovirus outbreak between October 2015 and September 2016. We associated in this dataset the monthly air travel volumes provided by the Colombian Civil Aviation Authority (AEROCIVIL) with the reported human cases of Zika Virus published by The Pan American Health Organization (PAHO). Our methodology consists of geocoding all the reported airports and identifying the catchment of each airport using the municipalities' boundaries since reported human cases of Zika Virus are available at the municipal level. In addition, we calculated the total population at risk in each airport catchment by combining the total population count in a catchment with the environmental suitability of the Aedes aegypti mosquito, the vector for the Zika virus. We separated the monthly air travel volumes into domestic and international based on the location of the origin airport. The current dataset includes the total air travel volumes of 23,539,364 passengers on domestic flights and 11,592,197 on international ones. We validated our dataset by comparing the monthly aggregated air travel volumes between airport catchments to those predicted by the gravity model. We hope the novel dataset will provide a resource to researchers studying the role of human mobility in the spread of mosquito-borne diseases and modeling disease spread in realistic networks.
\end{abstract}

% keywords can be removed

\section{Background}

Human movements have long been recognized as a major contributor to the spread of infectious diseases across the globe \cite{charniga2021epidemiology, findlater2018human}. This become more evident particularly during the past couple of decades because of the efficiency of travel networks and the increase of tourism and economic accessibility to air travel for a wider population \cite{findlater2018human}. Sources of data available for the study of human mobility vary between air travel records \cite{nah2016estimating}, internal migration \cite{sorichetta2016mapping}, GPS logs, phones, and microdata \cite{blumenstock2012inferring, garcia2015modeling}. Recent studies indicated that incorporating human mobility data into epidemiological studies has improved the modeling and analysis of the infectious spreading pathways \cite{kraemer2017spread}. However, curated human mobility datasets in underdeveloped countries with frequent outbreaks remain under-restricted use \cite{perrotta2022comparing} or lacking, which creates an obstacle for researchers \cite{wardle2023gaps}.

The recent introduction and spread of vector-borne diseases such as Dengue, chikungunya, and Zika virus in the Americas were largely attributed to human mobility because of the limited ability of the main vector, the Aedes aegypti and Ae. albopictus mosquitoes, to travel for longer distances. Although the vectors could be transported via the human transport networks and infect non-travelers \cite{findlater2018human}. Several human mobility data were developed in areas affected by vector-borne diseases such as the use of internal migration patterns as a proxy of human mobility in malaria-endemic countries including Colombia \cite{sorichetta2016mapping}. Further improvements were introduced to enhance the spatial resolution of internal migration data \cite{siraj2020modeling}. However, the coarse temporal resolution associated with these sources and their focus on long-term migration makes these sources disadvantageous to study and model the distribution of vector-borne diseases \cite{wardle2023gaps}. 
 
The curating of this dataset is motivated by the need to develop open-access datasets that integrate both human mobility and reported infectious disease cases at a spatial and temporal resolution adequate for informing predictive modeling and warning systems. In this dataset, Colombian air travel records were integrated with the reported human cases of Zika virus during the introduction of Zika virus in Colombia at a monthly temporal resolution. Airport catchments were defined using the administrative municipal boundaries to facilitate the integration of human-reported cases. The counts of the total population and total population at risk were estimated for each catchment. %The novel dataset will provide a benchmark for modeling, machine learning, and network analysis applications to study disease spreading. %

The paper is organized as follows: discussion of the different steps used to collect the data and define the catchments from airport codes (sections 2 to 4), estimating the population counts for each catchment (section 5), description of the final dataset (section 6), validation results (section 7) and a summary and future usage (section 8).   

\section{Data processing and preparation}

We used the official air flight dataset record to reconstruct the mobility network over Colombia. The raw data is provided by the Colombian government (aerocivil.gov.co) and provide the total aggregated number of passengers traveling between pairs of origin and destination airports. We geo-coded the unique airport codes in this dataset to identify the geographic location of each airport using Google Geocoding API. We assigned a geographic location to all airports in the trips data by initially matching using the IATA and local airport codes, to get coordinates (latitudes and longitudes) from the airports dataset. Our initial match resulted in 30,139 matches out of 32,332 trips. We then run Google’s Geocoder on the remaining 2,193 trips, to identify their coordinates using the trip city names.

\section{Identifying airports' catchments}

An approximation for airport catchment was generated using Voronoi tessellations applied to the airports’ coordinates. The approximate Voronoi catchment of each airport was then refined using the administrative municipalities' boundaries by identifying those municipalities that intersected with the Voronoi polygon of each airport. This step was important to allow for attaching the reported Zika virus cases, which are available at the municipal level. To avoid assigning the same municipality to more than one airport catchment (Voronoi tessellation), a population-weighted strategy was adopted. The population-weighted strategy is based on identifying the shared population between each airport catchment (Voronoi polygon) and each intersecting municipality with that catchment. The shared population at the intersection was estimated using the Worldpop unconstrained and UN-adjusted raster of population counts raster at a spatial resolution of 100 m \cite{tatem2017worldpop}. Municipalities were then assigned to a single airport catchment based on the largest shared population from all intersecting airport catchments.

\begin{figure}
\centering
\includegraphics[width=7cm]{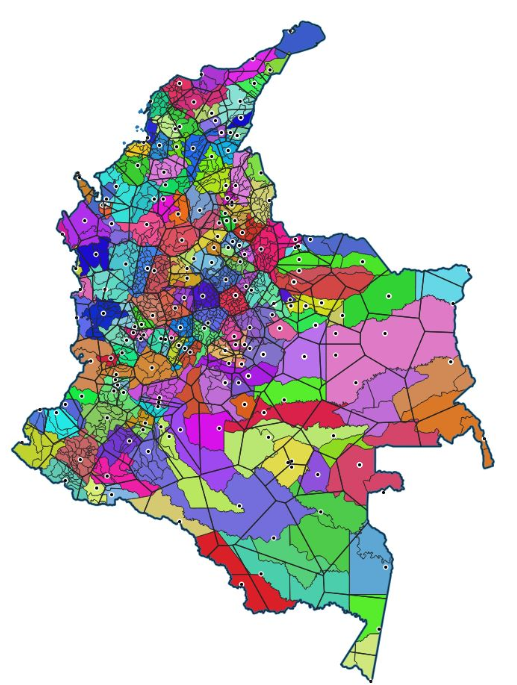}
\caption{Geographic mismatches between administrative Municipal boundaries in color and Voronoi approximations of airport catchments (straight black lines). Mismatches in the eastern part of Colombia are the result of finding multiple airports within the same municipality and vice versa in the western part of the country.}
\label{catchment}
\end{figure}

Various mismatches were observed between the airports' catchments and municipalities’ boundaries, which can be seen in Figure \ref{catchment}. These mismatches are mainly attributed to the difference in size between the catchments (Voronoi polygons) and municipalities. For example, in the eastern side of Colombia, some municipalities include more than a single airport. While on the western side, the size of municipalities is relatively smaller and a single airport catchment could contain multiple municipalities. In order to account for these mismatches, multiple airports that fall within a single municipality were aggregated together as a single catchment defined using the municipality boundaries, while multiple municipalities falling within a single airport catchment were dissolved into a single catchment.  

The number of airports changed between years due to the addition and closure of smaller airports over time, which were identified by the smaller number of travelers and validated visually using google maps. We stabilized the number of airports between 2013 - 2016 by combining the limited number of travelers from the smaller airports to the nearest larger airport. This step was conducted to avoid changes in the geometry of the catchments over time and construct a temporal network.  

\section{Integrating  Air Travel flows and human reported cases }%to each airport catchment}

Air travel volumes were aggregated based on the catchments on monthly steps. The catchments are defined using the municipalities as described in section 3. Each record in the dataset provides the total number of passengers traveled from an origin catchment to a destination catchment within a calendar month. The dataset is designed to construct a weighted network for air travel where the catchments are nodes and the origin-destination pairs are the edges weighted by the number of travelers within a month.

Defining catchments using municipal boundaries allowed for attaching the number of reported human cases to each catchment. If multiple catchments are aggregated because they fall within the area of a single airport, then their reported human cases were combined. Human cases are reported weekly but aggregated into monthly steps to match the temporal granularity of the travel data. The monthly steps provide an adequate period for the Zika virus and an improvement over other existing datasets such as internal migration \cite{sorichetta2016mapping} given that the Zika virus incubation period is 3-14 days \cite{krow2017estimated} and the median duration of infectiousness reported to be 39.6 days \cite{counotte2018sexual}.   

\section{Estimating total population at risk }
\begin{figure}
\centering
\includegraphics[width=14cm]{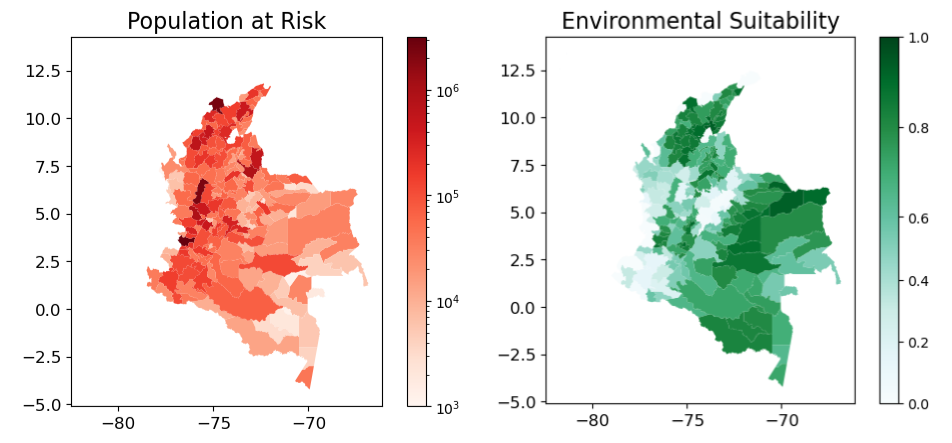}
\caption{ The spatial distribution of total population at risk of contracting Zika Virus and the Environmental suitability for the Aedes aegypti across airport catchments}
\label{environ_pop}
\end{figure}

\begin{figure}
\centering
\includegraphics[width=14cm]{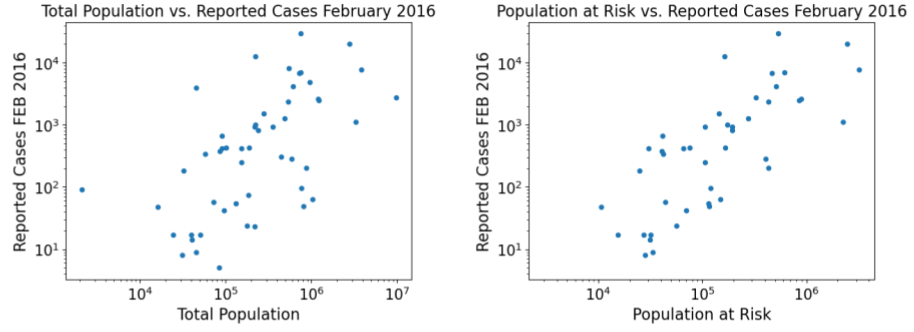}
\caption{The concordance between reported Zika virus cases per catchment and total population at risk is stronger when compared to concordance with the total population.}
\label{corr}
\end{figure}

In order to support modeling and the study of the role of human mobility in the transmission of the vector-borne outbreak, the total population at risk of contracting the Zika virus was estimated for each airport catchment. The total population at risk was calculated by taking the cross product of the total population count extracted from the WorldPop unconstrained UN-adjusted 100-meter rasters (\url{https://hub.worldpop.org/}) and the environmental suitability for the Aedes aegypti mosquitoes available at 5 km resolution \cite{kraemer2015global}. The population count was aggregated to 5 km before taking the cross-product in order to match the spatial resolution of the environmental suitability raster. Both environmental suitability and total population at risk showed spatial heterogeneity in Figure \ref{environ_pop}. The comparison between the number of reported Zika virus cases per airport catchment before and after correcting for environmental suitability (Figure \ref{corr}) indicates that the association of reported cases and the population is stronger when considering only the population at risk compared to the total population (Pearson correlation coefficient of 0.443 and 0.229 respectively). It is important to mention that the estimation of the total population at risk does not account for the number of recovered  cases from previous time periods. This should be considered in the case of catchments with low populations more than highly populated ones, where the ratio of recovered to total population at risk is small.   

\section{Description of Data} 
 % tabular data fields 
 %  \subitem integrated monthly flows and cases by catchment
 %  \subsubitem domestic
 %  \subsubitem international 
 %  \subitem aggregation of municipalities to catchments
 %  \item  geographic datasets airport catchment 
 %  \subitem description of the geographic units 
 %  \subitem attributes of units 
 %  \subsubitem total population
 %  \subsubitem total population at risk 
 %  \subsubitem  environmental suitability for adaes

The data is available through the Illinois Data Bank \cite{stein2018meaningful} and contains the following files: 

\begin{itemize}
  \item \textbf{Airport catchment:} a geojson file projected using (EPSG: 6933, EASE-Grid 2.0) and contains the following fields: geometry of the catchment polygons, \texttt{Muni\_ID}: catchment unique id, \texttt{pop\_risk\_kraemer\_sum}: total count of the population at risk, \texttt{total\_population\_sum}: total count of the population, and suitability: average environmental suitability for the Aedes aegypti mosquito of each catchment.  
 
\item \textbf{Air Travel Flows:} CSV files with each row representing the total travelers' volume between an origin (o) and destination (d) catchments during a given month. In addition, the dataset provides the total numbers of reported cases per month at the origin and destination catchments separately with different degrees of confidence designated from the least to the most confidence level as suspected (S), clinic-suspected (SC), clinic-confirmed (CC), lab-confirmed (CL) and a total of all reported cases (sum).  Air travel flows are separated into domestic (between catchments in Colombia) and international (travel between an international country and an airport catchment in Colombia). The classification was based if the origin or the destination airport is within or outside the boundaries of Colombia. In the case of the international dataset, countries are treated as a single catchment that exchanges travelers with domestic catchment within Colombia. 
 
\item \textbf{Aggregated municipality dictionary:} a CSV file defining the correspondence between the official Colombian government municipality code (CCNCT) and the airport catchment id (\texttt{Aggregate\_ID}). This file is useful in identifying how municipalities were assigned to different airport catchments.

\item \textbf{Airport municipality dictionary}: CSV file identifying the correspondence between airports identified by airport IATA code and airport catchments id. We combined both the domestic and international airports in the same list, where international airports are designated with catchment id above 205. Multiple airports belong to the same catchment if this airport falls within a single municipality.  
 
 \item \textbf{Nation ID Dictionary:} a CSV file that has the corresponding country name (\texttt{Nation\_Name}) and country unique id (\texttt{Nation\_Code}). The country unique id is identical to the airport catchment id in the Airport municipality dictionary since countries are treated as a single catchment with ids above 205.

\end{itemize}

\section{Data Validation} 

\begin{figure}
\centering
\includegraphics[width=9cm]{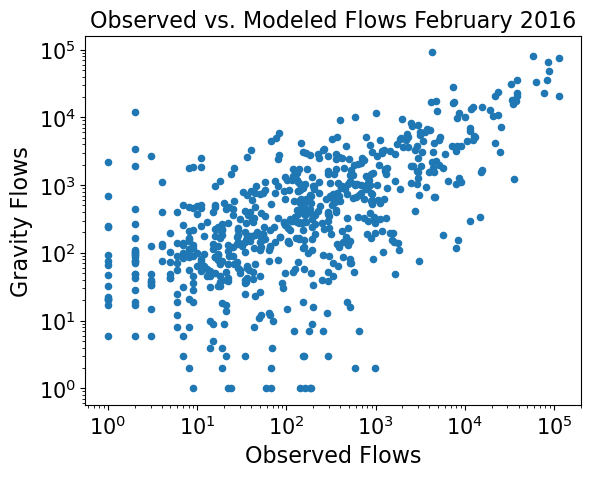}
\caption{Comparison between observed flows and predicted using the gravity model. }
\label{gravity}
\end{figure}

The aggregated flows from the Air travel records were validated by comparing them to the domestic flows predicted using the gravity model. The gravity models implementation in sci-kit mobility \cite{pappalardo2019scikit} requires the size of populations at both the origin and destination nodes and the geographic distance between them to produce the percentage of travelers from a given origin to a given destination as probabilities relative to total travel from that origin. The predicted flow volumes of travelers for each month were calculated using the cross-product of the gravity model probabilities and observed monthly total number of travelers from each catchment. The comparison of observed and predicted flows shows a great alignment (Pearson correlation coefficient of 0.74). as shown in Figure \ref{gravity}. There are two groups of catchments that the gravity model under and overestimates the volumes of travelers because the gravity model does not include the attractiveness and retention of a region. For example, the capital Bogotá is a major attraction for domestic travelers and produces a limited number of travelers proportional to its population due to high retention, while other catchments might be repulsive and produce more travelers compared to what is predicted by the gravity model.  

\section{Summary}

We presented a novel open dataset that integrates monthly air travel volumes with reported human cases of the Zika virus in Colombia during the Zika outbreak of 2015-2016. We overcome the challenges of linking these two data sources by defining consistent airport catchments using the municipal boundaries, which allowed for attaching the reported Zika virus cases. In addition, the total population at risk of contracting the Zika virus was estimated for each catchment. This dataset fills in the gap of accessing human mobility datasets that are at an adequate spatial and temporal resolution to study the introduction of vector-borne disease in the Americas, especially since other alternatives are either too coarse (i.e., internal migration) or strictly restricted from public use (i.e., Phone records). In addition, this dataset provides an extended estimate of travel volumes outside the Zika outbreak window between the years 2013-2020, which will help researchers interested in modeling other vector-borne outbreaks, such as the Chikungunya outbreak. The aim of publishing this dataset is to support researchers in the domains of disease ecology, network science, and graph machine learning.

\section{Data Availability}
The Dataset is available through this Github
link: \url{https://github.com/a2soliman/Air_Travel_Colombia} and will be available upon publication through the Illinois Data Bank. 

\section*{Acknowledgment}
This work is supported by funding from the National Science Foundation, award number 1824961.

\bibliographystyle{unsrtnat}
\bibliography{references}  %%% Uncomment this line and comment out the ``thebibliography'' section below to use the external .bib file (using bibtex) .

\begin{thebibliography}{16}
\providecommand{\natexlab}[1]{#1}
\providecommand{\url}[1]{\texttt{#1}}
\expandafter\ifx\csname urlstyle\endcsname\relax
  \providecommand{\doi}[1]{doi: #1}\else
  \providecommand{\doi}{doi: \begingroup \urlstyle{rm}\Url}\fi

\bibitem[Charniga(2021)]{charniga2021epidemiology}
Kelly Charniga.
\newblock Epidemiology of the 2014-2017 zika and chikungunya epidemics in
  colombia.
\newblock 2021.

\bibitem[Findlater and Bogoch(2018)]{findlater2018human}
Aidan Findlater and Isaac~I Bogoch.
\newblock Human mobility and the global spread of infectious diseases: a focus
  on air travel.
\newblock \emph{Trends in parasitology}, 34\penalty0 (9):\penalty0 772--783,
  2018.

\bibitem[Nah et~al.(2016)Nah, Mizumoto, Miyamatsu, Yasuda, Kinoshita, and
  Nishiura]{nah2016estimating}
Kyeongah Nah, Kenji Mizumoto, Yuichiro Miyamatsu, Yohei Yasuda, Ryo Kinoshita,
  and Hiroshi Nishiura.
\newblock Estimating risks of importation and local transmission of zika virus
  infection.
\newblock \emph{PeerJ}, 4:\penalty0 e1904, 2016.

\bibitem[Sorichetta et~al.(2016)Sorichetta, Bird, Ruktanonchai,
  zu~Erbach-Schoenberg, Pezzulo, Tejedor, Waldock, Sadler, Garcia, Sedda,
  et~al.]{sorichetta2016mapping}
Alessandro Sorichetta, Tom~J Bird, Nick~W Ruktanonchai, Elisabeth
  zu~Erbach-Schoenberg, Carla Pezzulo, Natalia Tejedor, Ian~C Waldock, Jason~D
  Sadler, Andres~J Garcia, Luigi Sedda, et~al.
\newblock Mapping internal connectivity through human migration in malaria
  endemic countries.
\newblock \emph{Scientific data}, 3\penalty0 (1):\penalty0 1--16, 2016.

\bibitem[Blumenstock(2012)]{blumenstock2012inferring}
Joshua~E Blumenstock.
\newblock Inferring patterns of internal migration from mobile phone call
  records: evidence from rwanda.
\newblock \emph{Information Technology for Development}, 18\penalty0
  (2):\penalty0 107--125, 2012.

\bibitem[Garcia et~al.(2015)Garcia, Pindolia, Lopiano, and
  Tatem]{garcia2015modeling}
Andres~J Garcia, Deepa~K Pindolia, Kenneth~K Lopiano, and Andrew~J Tatem.
\newblock Modeling internal migration flows in sub-saharan africa using census
  microdata.
\newblock \emph{Migration Studies}, 3\penalty0 (1):\penalty0 89--110, 2015.

\bibitem[Kraemer et~al.(2017)Kraemer, Faria, Reiner, Golding, Nikolay, Stasse,
  Johansson, Salje, Faye, Wint, et~al.]{kraemer2017spread}
Moritz~UG Kraemer, Nuno~R Faria, Robert~C Reiner, Nick Golding, Birgit Nikolay,
  Stephanie Stasse, Michael~A Johansson, Henrik Salje, Ousmane Faye, GR~William
  Wint, et~al.
\newblock Spread of yellow fever virus outbreak in angola and the democratic
  republic of the congo 2015--16: a modelling study.
\newblock \emph{The Lancet infectious diseases}, 17\penalty0 (3):\penalty0
  330--338, 2017.

\bibitem[Perrotta et~al.(2022)Perrotta, Frias-Martinez, Pastore~y Piontti,
  Zhang, Luengo-Oroz, Paolotti, Tizzoni, and Vespignani]{perrotta2022comparing}
Daniela Perrotta, Enrique Frias-Martinez, Ana Pastore~y Piontti, Qian Zhang,
  Miguel Luengo-Oroz, Daniela Paolotti, Michele Tizzoni, and Alessandro
  Vespignani.
\newblock Comparing sources of mobility for modelling the epidemic spread of
  zika virus in colombia.
\newblock \emph{PLoS Neglected Tropical Diseases}, 16\penalty0 (7):\penalty0
  e0010565, 2022.

\bibitem[Wardle et~al.(2023)Wardle, Bhatia, Kraemer, Nouvellet, and
  Cori]{wardle2023gaps}
Jack Wardle, Sangeeta Bhatia, Moritz~UG Kraemer, Pierre Nouvellet, and Anne
  Cori.
\newblock Gaps in mobility data and implications for modelling epidemic spread:
  a scoping review and simulation study.
\newblock \emph{Epidemics}, page 100666, 2023.

\bibitem[Siraj et~al.(2020)Siraj, Sorichetta, Espa{\~n}a, Tatem, and
  Perkins]{siraj2020modeling}
Amir~S Siraj, Alessandro Sorichetta, Guido Espa{\~n}a, Andrew~J Tatem, and
  T~Alex Perkins.
\newblock Modeling human migration across spatial scales in colombia.
\newblock \emph{Plos one}, 15\penalty0 (5):\penalty0 e0232702, 2020.

\bibitem[Tatem(2017)]{tatem2017worldpop}
Andrew~J Tatem.
\newblock Worldpop, open data for spatial demography.
\newblock \emph{Scientific data}, 4\penalty0 (1):\penalty0 1--4, 2017.

\bibitem[Krow-Lucal et~al.(2017)Krow-Lucal, Biggerstaff, and
  Staples]{krow2017estimated}
Elisabeth~R Krow-Lucal, Brad~J Biggerstaff, and J~Erin Staples.
\newblock Estimated incubation period for zika virus disease.
\newblock \emph{Emerging infectious diseases}, 23\penalty0 (5):\penalty0 841,
  2017.

\bibitem[Counotte et~al.(2018)Counotte, Kim, Wang, Bernstein, Deal, Broutet,
  and Low]{counotte2018sexual}
Michel~Jacques Counotte, Caron~Rahn Kim, Jingying Wang, Kyle Bernstein,
  Carolyn~D Deal, Nathalie Jeanne~Nicole Broutet, and Nicola Low.
\newblock Sexual transmission of zika virus and other flaviviruses: a living
  systematic review.
\newblock \emph{PLoS medicine}, 15\penalty0 (7):\penalty0 e1002611, 2018.

\bibitem[Kraemer et~al.(2015)Kraemer, Sinka, Duda, Mylne, Shearer, Barker,
  Moore, Carvalho, Coelho, Van~Bortel, et~al.]{kraemer2015global}
Moritz~UG Kraemer, Marianne~E Sinka, Kirsten~A Duda, Adrian~QN Mylne, Freya~M
  Shearer, Christopher~M Barker, Chester~G Moore, Roberta~G Carvalho,
  Giovanini~E Coelho, Wim Van~Bortel, et~al.
\newblock The global distribution of the arbovirus vectors aedes aegypti and
  ae. albopictus.
\newblock \emph{elife}, 4:\penalty0 e08347, 2015.

\bibitem[Stein and Dunham(2018)]{stein2018meaningful}
Ayla Stein and Elise Dunham.
\newblock Meaningful data sharing: Developing the illinois data bank metadata
  framework.
\newblock \emph{Journal of Library Metadata}, 18\penalty0 (2):\penalty0 59--83,
  2018.

\bibitem[Pappalardo et~al.(2019)Pappalardo, Simini, Barlacchi, and
  Pellungrini]{pappalardo2019scikit}
Luca Pappalardo, Filippo Simini, Gianni Barlacchi, and Roberto Pellungrini.
\newblock scikit-mobility: A python library for the analysis, generation and
  risk assessment of mobility data.
\newblock \emph{arXiv preprint arXiv:1907.07062}, 2019.

\end{thebibliography}

%%% Uncomment this section and comment out the \bibliography{references} line above to use inline references.
% \begin{thebibliography}{1}

% 	\bibitem{kour2014real}
% 	George Kour and Raid Saabne.
% 	\newblock Real-time segmentation of on-line handwritten arabic script.
% 	\newblock In {\em Frontiers in Handwriting Recognition (ICFHR), 2014 14th
% 			International Conference on}, pages 417--422. IEEE, 2014.

% 	\bibitem{kour2014fast}
% 	George Kour and Raid Saabne.
% 	\newblock Fast classification of handwritten on-line arabic characters.
% 	\newblock In {\em Soft Computing and Pattern Recognition (SoCPaR), 2014 6th
% 			International Conference of}, pages 312--318. IEEE, 2014.

% 	\bibitem{hadash2018estimate}
% 	Guy Hadash, Einat Kermany, Boaz Carmeli, Ofer Lavi, George Kour, and Alon
% 	Jacovi.
% 	\newblock Estimate and replace: A novel approach to integrating deep neural
% 	networks with existing applications.
% 	\newblock {\em arXiv preprint arXiv:1804.09028}, 2018.

% \end{thebibliography}

\end{document}